\documentclass[aps,prl,reprint,superscriptaddress]{revtex4-1}
\usepackage{graphicx,amsmath}
\usepackage{natbib}
\usepackage{multibib}
\usepackage{color}
\usepackage{scalerel}
\usepackage{times}
\usepackage{hyperref}
\hypersetup{colorlinks,urlcolor=blue,citecolor=blue,linkcolor=blue}

\begin{document}
\author{Zachary G. Nicolaou}
\affiliation{Department of Physics and Astronomy, Northwestern University, Evanston, Illinois 60208, USA}
\author{Hermann Riecke}
\affiliation{Department of Engineering Sciences and Applied Mathematics, Northwestern University, Evanston, Illinois 60208, USA}
\affiliation{Northwestern Institute on Complex Systems, Northwestern University, Evanston, Illinois 60208, USA}
\author{Adilson E. Motter}
\affiliation{Department of Physics and Astronomy, Northwestern University, Evanston, Illinois 60208, USA}
\affiliation{Northwestern Institute on Complex Systems, Northwestern University, Evanston, Illinois 60208, USA}
\title{Chimera States in Continuous Media: Existence and Distinctness}
\begin{abstract}
The defining property of chimera states is the coexistence of coherent and incoherent domains in systems that are structurally and spatially homogeneous. The recent realization that such states might be common in oscillator networks raises the question of whether an analogous phenomenon can occur in continuous media.  Here, we show that chimera states can exist in continuous systems even when the coupling is strictly local, as in many fluid and pattern forming media. Using the complex Ginzburg-Landau equation as a model system, we characterize chimera states consisting of a coherent domain of a frozen spiral structure and an incoherent domain of amplitude turbulence. We show that in this case, in contrast with discrete network systems, fluctuations in the local coupling field play a crucial role in limiting the coherent regions. We suggest these findings shed light on new possible forms of coexisting order and disorder in fluid systems.
\end{abstract}

\onecolumngrid\hfill 
{\small {\it Phys. Rev. Lett.} {\bf 119}, 244101 (2017) 

\hfill DOI: \href{https://doi.org/10.1103/PhysRevLett.119.244101}{10.1103/PhysRevLett.119.244101}\hspace{-1mm}}
\bigskip\twocolumngrid

\maketitle

Chimera states are spatiotemporal patterns resulting from symmetry breaking.
The discovery of such states in oscillator networks demonstrated that 
even in systems of identically-coupled identical oscillators, mutually synchronized oscillators 
can coexist with desynchronized ones \cite{Kuramoto_Battogtokh:2002}.
This coexistence is particularly remarkable because the coherent and incoherent domains are bidirectionally 
coupled: it is counterintuitive that the state would be persistent despite the perturbations
that desynchronized oscillators unavoidably  exert on synchronized ones, and vice versa. 
Chimera states were initially identified in networks of phase oscillators with nonlocal
coupling \cite{Kuramoto_Battogtokh:2002,Abrams_Strogatz:2004}, but they have since been demonstrated for
a wide range of oscillator networks \cite{Panaggio_Abrams:2015}.
Examples include networks with couplings that have delays  \cite{2010_Sheeba_Lakshmanan, 2013_Larger_Maistrenko},
have time dependence or noise \cite{2015_Buscarino_Hovel, 2016_Loos_Zakharora}, and are global \cite{2014_Yeldesbay_Roseblum,2014_Sethia_Sen} or local  \cite{2015_Laing,2016_Bera_Ghosh,2016_Clerc_Rojas,2016_Li_Dierckx}; 
they also include networks of  phase-amplitude oscillators  \cite{2010_Bordyugov_Rosenblum,2013_Sethia_Johnston}, inertial oscillators \cite{2014_Bountis_Bezerianos,2015_Olmi}, and chaotic oscillators \cite{2011_Omelchenko_Scholl,2013_Gu_Davidsen}.
 Moreover, chimera states have been  observed experimentally in various systems, including networks of
 optical \cite{2012_Hagerstrom_Scholl}, chemical  \cite{2012_Tinsley_Kenneth},
  and mechanical \cite{2013_Martens_Hallatschek} oscillators.
Yet, with very few exceptions \cite{2014_Skardal_Restrepo,2014_Schmidt_Morales},  
previous work has focused exclusively on chimeras in (discrete) network  
systems. It is thus natural to ask the extent to which chimeras states can exist and have salient properties in continuous  systems.

We first note that continuous systems can exhibit analogous examples of coexisting order and disorder in homogenous
 media,  but the connection between those phenomena and chimera states has remained largely unappreciated.  
 Perhaps the most significant examples occur in fluid mechanics.  Consider, for instance, a Taylor-Couette flow, where 
 the fluid  is constrained to the space between two rotating cylinders.  As the rate of rotation increases the dynamics 
change from an orderly laminar regime to a turbulent one through a series of intermediate dynamical states \cite{Andereck_Harry:1986}, 
including a spiral turbulence flow regime characterized by a persistent spiraling region of turbulent flow that 
coexists with a domain of laminar flow \cite{Hegseth_Pomeau:1989}. The spiral turbulence in this system is 
thus a candidate for a fluid counterpart of a chimera state. Related examples can be found in parametrically forced 
Faraday waves \cite{1996_Kudrolli_Gollub, Granzow_Riecke:1996}, where fluid driven by an oscillating support
can form coexisting domains of regular stripes and chaotic surface waves, and in the spatiotemporal intermittency regime of Rayleigh-B\'enard convection \cite{1988_Ciliberto_Bigazzi}.
A key difference in continuous systems is in the nature of the coupling, 
which often consists of a strictly local (differential) component that acts in the limit of small spatial scales.
Modeling chimera states in continuous systems and establishing that they can exist in the absence of  any nonlocal coupling 
thus remains an important outstanding problem in this field. Although working directly with fluid equations 
is possible in principle, to address this problem it is more enlightening to employ simpler model equations.

In this Letter, we report on chimera states in the locally-coupled complex Ginzburg-Landau (CGL) equation in two spatial dimensions.  These states, which we refer to as {\em frozen vortex chimeras}, correspond to coexisting domains of frozen spirals and amplitude turbulence and are characterized in a previously underexplored parameter regime of the system. 
They are distinct from spiral wave chimeras 
previously identified in discrete systems \cite{2010_Martens_Laing_Strogatz, 2012_Omelchenko_Olesandr,2015_Pannagio_Abrams2}  
in that the core of the frozen spirals is coherent and the media lose coherence far from the core, and not the other way around. 
We analyze these states by introducing a  local coupling field generalization of the  Kuramoto-Battogtokh approach \cite{Kuramoto_Battogtokh:2002}. 
Crucially, we show that fluctuations in the coupling field cannot be neglected for such locally-coupled continuous systems, which set them fundamentally apart from previously studied network analogs.  

\begin{figure}[b]
\includegraphics[width=\columnwidth]{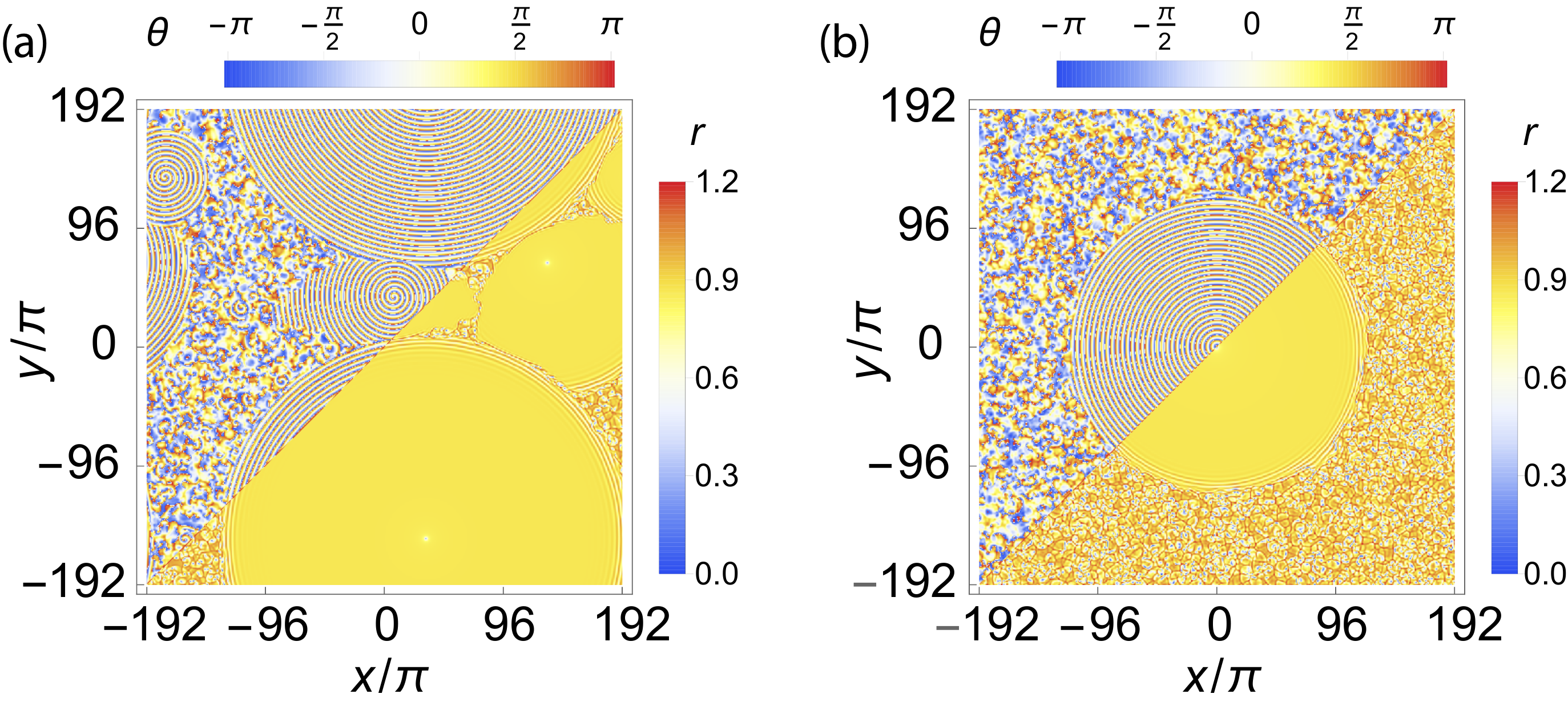}
\caption{ 
Coexistence of coherent spirals and amplitude turbulence in the CGL system \eqref{cgle}: (a) with 
$c_1=1.5$ and $c_3=0.77$
and random initial conditions after a time  
$t=10^4$; (b) with 
$c_1=2.0$ and $c_3=0.85$ and a spiral initial condition after a time 
$t=10^4$.   The phase $\theta\equiv\arg(A)$ is depicted in the upper left and the amplitude $r\equiv|A|$ in the lower right. \label{fig0}}
\end{figure}

 In the study of pattern formation, 
 the CGL equation
\begin{equation}
\label{cgle}
\frac{\partial A}{\partial t} = A + (1+ic_1)\nabla^2 A - (1-ic_3)|A|^2A
\end{equation}
describes the universal behavior of a homogenous oscillatory medium in the vicinity of a supercritical Hopf bifurcation, where $A=A(x,y;t)$ is a complex-valued field. Modeling applications of this equation include examples of
Rayleigh-B\'enard convection \cite{Cross_Hohenberg:1993,1997_Liu_Ecke,Madruga_Werner:2006} and the Belousov-Zhabotinsky (BZ) reaction \cite{1994_Kramer_Walgraef,1996_Ouyang_Flesselles}.
The CGL system can exhibit a variety of dynamical phases depending on the parameters $c_1$ and $c_3$  \cite{Aranson_Kramer:2002}.  
As in other nonlinear wave systems \cite{1996_Campbell}, these phases include different coherent and localized structures. 
The most disordered phase is that of amplitude or defect turbulence, with a disordered and finite density of defects where $|A|$ reaches zero (and the phase of $A$ is undefined) at a point.  
A second important phase consists of frozen spiral structures, where $|A|$ becomes time independent near the spiral core
and the phase of $A$ has periodic spiraling structures.  In particular, states with slowly evolving domains of frozen spirals, so-called vortex glass states, have attracted significant attention \cite{Huber_Bohr:1992,1992_Aranson_Weber,Brito_Chate:2003,Kevrekidis_Rasmussen:2001}.  
One relevant parameter regime previously studied corresponds to $c_1=2.0$ and  $c_3<0.75$,
in which the frozen (anti)spirals \cite{antispirals} can nucleate out of amplitude turbulence and grow to a limited size \cite{Chate_Manneville:1996}.
 Of special interest to our research question would be a parameter regime that supports frozen spiral states in a turbulent sea but with no growth and no spiral nucleation over sufficiently long time scales.

Figure \ref{fig0} shows results of our simulations of the CGL system \eqref{cgle} as a function of the spatial coordinates $x$ and $y$. A regime in which coherent spirals have nucleated out of the amplitude turbulence and grown to their maximum size, with interstitial patches of residual amplitude turbulence between them, is shown in Fig.\,\ref{fig0}(a). 
As $c_1$ and $c_3$ are increased, on the other hand, the average time $T_{\mathrm{nuc}}$ required for spiral nucleation (in a simulation area $L^2$ starting from an initial state of full amplitude) turbulence rises sharply. For 
$c_1=2.0$ and $c_3=0.85$, for example,
no spirals nucleated 
out of amplitude turbulence for times up to $t=10^6$ in ten simulations (with independent initial conditions); below we estimate that the nucleation time is in fact $10^3$ times longer than that.  However, spirals and amplitude turbulence can still coexist in this new regime;
Fig.\,\ref{fig0}(b) shows a frozen vortex chimera after 
$t=10^4$, which was obtained with an initial condition consisting of a single spiral. This initial condition was generated starting with a smaller value of $c_3$ for which nucleation occurs rapidly and then adiabatically increasing $c_3$ after one spiral had nucleated.
These numerical simulations are implemented using accurate discrete approximations of the continuous CGL system of interest.  In all simulations the system is taken to have linear size  $L=384\pi$ and is integrated using a pseudospectral algorithm with $N=1536$ modes in each dimension \cite{github}.  We carried out a detailed study of these results with increasingly finer spatial grids and time steps (see Supplemental Material, Sec.\,S1 \cite{SM}).  Crucially, we employ everywhere sufficiently fine spatial grids and time steps to ensure convergence to the continuum limit.
Thus, these long-lived spatiotemporal patterns are {\em continuous-media chimera states} exhibiting a sizable coherent region (the spiral) and a sizable incoherent region (the amplitude turbulence phase).

\begin{figure}[b]
\includegraphics[width=1.0\columnwidth]{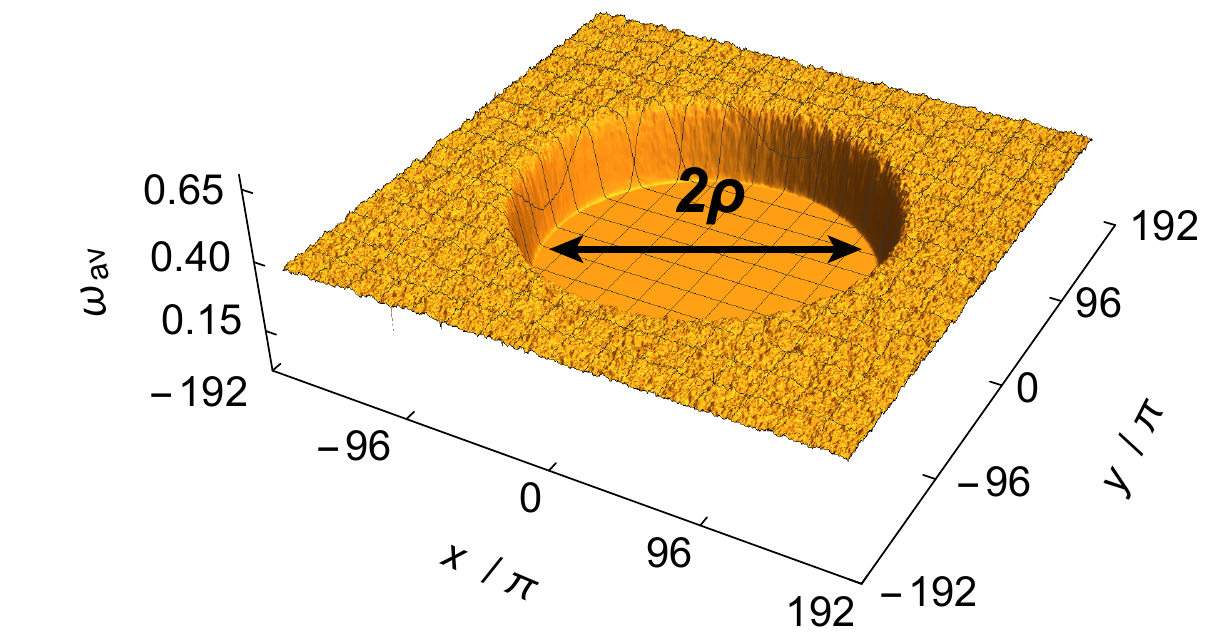}
\caption{
Time-averaged angular frequency 
$\omega_{\mathrm{av}}\equiv\langle\frac{d\theta}{dt}\rangle$
for the frozen vortex chimera in Fig.\,\ref{fig0}(b), where the arrow indicates the diameter $2\rho$ of the coherent region.
The mean frequency in the coherent region is $\Omega=0.15$, while the frequency is significantly higher 
($>0.40$) in the incoherent region. The distinction between the regions does not depend sensitively on the averaging time interval (here taken to be $10^4$).  For an animation of the time evolution of this chimera state, see Supplemental Material \cite{SM}.
 \label{fig2}}
\end{figure} 

\begin{figure}[t]
\includegraphics[width=\columnwidth]{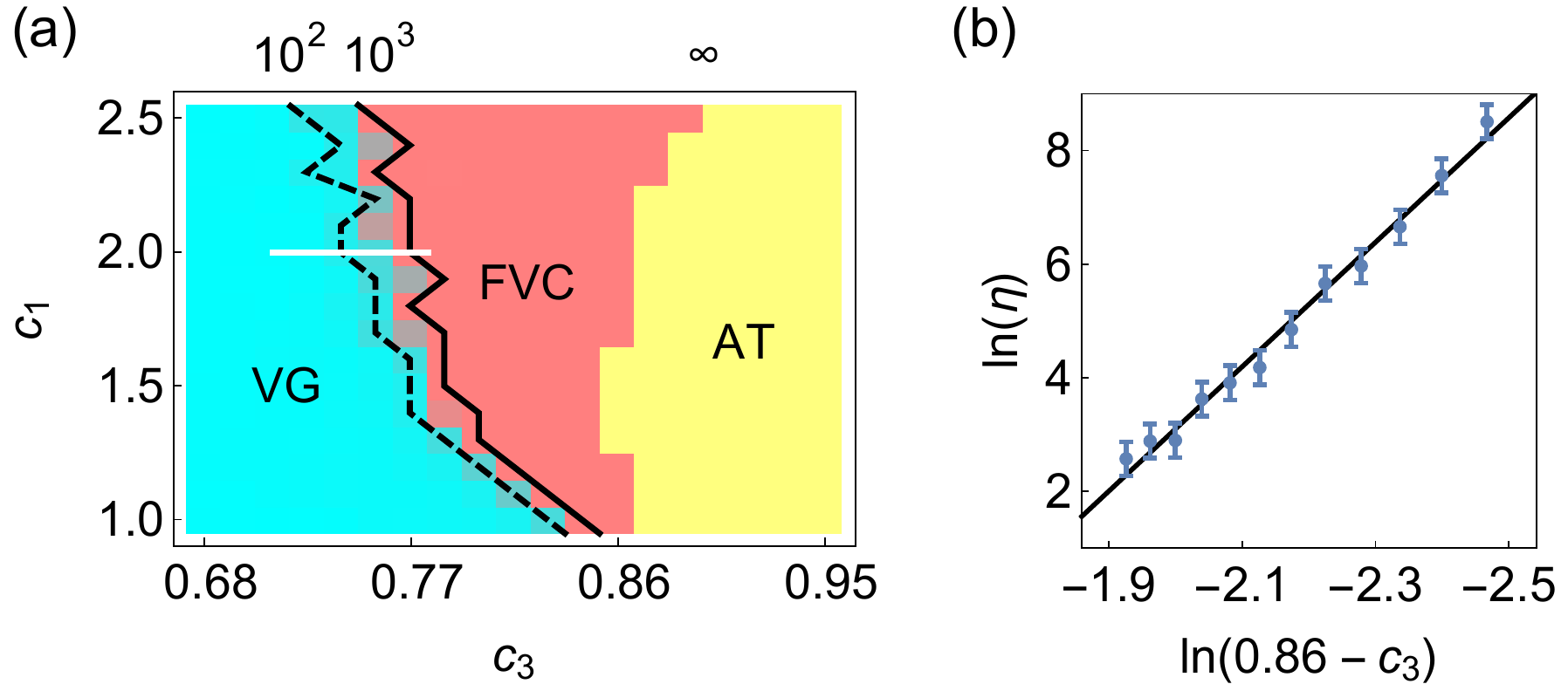}
\caption{ (a) Diagram of dynamical phases of the CGL system, where the frozen vortex chimeras (FVC), vortex glass (VG), and amplitude turbulence (AT) occupy the intermediate, small, and large $c_3$ regions, respectively.
The boundary between VG and FVC was determined by calculating $\eta$ from five realizations of spiral nucleation from an amplitude turbulence initial state;
the black lines show a modest change in this boundary as the threshold $\eta$  
changes from $10^2$ (dashed line) to $10^3$ (solid line).  
The boundary between FVC and AT was determined by adiabatically varying $c_1$ and $c_3$ to move a spiral initial condition past the point of spiral destabilization, where $\eta$ is predicted to diverge.
(b) Log-log plot of $\eta$
for $c_1=2.0$ as a function of $0.86-c_3$, corresponding to the white line in (a). The error bars show the estimated standard deviation, and the line marks the linear fit. 
\label{fig4}}
\end{figure}

Figure \ref{fig2} shows the time-averaged frequency as a function of position in a frozen vortex chimera with $c_1=2.0$ and 
$c_3=0.85$.
At the center there is a coherent domain, which is a frozen spiral of radius $\rho$ with low mean angular frequency $\Omega$, 
whereas the outer domain is occupied by an incoherent region, which has higher mean frequency and exhibits amplitude turbulence. The quantities $\rho$ and $\Omega$ establish the natural length and time scales of the spiral.
We then formally define a frozen vortex chimera as a state in which there exists a spiral (of area $\pi\rho^2$) and a surrounding neighborhood of amplitude turbulence 
of area $O(\pi\rho^2)$ that
persists without appreciable change for a time interval much longer (i.e., orders of magnitude larger) than the spiral oscillation period of $2\pi/\Omega$. Multi-spiral frozen vortex chimeras can be defined analogously.

We now consider the parameter range over $c_1$ and $c_3$ in which such frozen vortex chimeras exist. These chimeras are intermediate states between the vortex glass phase and the amplitude turbulence phase, as shown in Fig.\,\ref{fig4}(a). 
We first note that if a spiral is to coexist with amplitude turbulence for many periods of oscillation and qualify as a chimera, then (i)  the spiral must persists in its environment (which sets the boundary with the amplitude turbulence phase) and (ii) the rate of spiral nucleation in its neighborhood must be small compared to its angular frequency $\Omega$ (which sets the boundary with the vortex glass phase).  
To quantify the vortex glass transition, we define $\eta\equiv(L^2/\pi \rho^2)[T_{\mathrm{nuc}}/ (2\pi/\Omega)]$, which is the nucleation time $T_{\mathrm{nuc}}$ properly normalized by the spiral period $2\pi/\Omega$ and the normalized neighborhood area $\pi\rho^2/L^2$.   
Our definition of a frozen vortex chimera requires $\eta$ to be orders of magnitude larger than $1$, while states with faster spiral nucleation, and hence smaller $\eta$, are considered a vortex glass.
Systematic simulations along the line $c_1=2.0$ revealed that frozen vortex chimeras exist up to values of $c_3\lesssim0.86$.
Figure \ref{fig4}(b) shows a log-log plot of the normalized nucleation time $\eta$ versus $0.86-c_3$ for $c_1=2.0$. 
The transition boundary between vortex glasses and frozen vortex chimeras thus does not depend sensitively on the threshold for $\eta$, since the increase in $\eta$ is extremely stiff as $c_3$ is increased, as demonstrated by the contour lines in Fig.\,\ref{fig4}(a) and the steep slope (with scaling exponent of approximately $-10$) in Fig.\,\ref{fig4}(b).  Like other chimera states \cite{Wolfrum_Omelchenko:2011}, frozen vortex chimeras are transient states. The mechanism of their collapse into coherent states is through the nucleation of new spirals in the neighborhood of the coherent domain. This process can take an exceedingly long time. For example, the scaling in Fig.\,\ref{fig4}(b) indicates that the lifetime of the frozen vortex chimera in Fig.\,\ref{fig2} is over a billion spiral oscillation periods.

It follows from Fig.\,\ref{fig4}(a) that the parameter regime where frozen vortex chimeras prevail is relevant for experimentally accessible systems such as the BZ reaction system  \cite{1994_Kramer_Walgraef,1996_Ouyang_Flesselles}.  In models of the BZ system, the CGL parameter $c_1$ is determined primarily by species diffusion coefficients, while the parameter $c_3$ is determined by reaction rates and concentrations. By varying sulfuric acid concentrations, for example, the parameter regime $c_1=1.4$ and $0.5<c_3<0.7$ just below the transition between vortex glasses and frozen vortex chimeras (around $c_3\approx0.8$ in Fig.\,\ref{fig4}) has been experimentally explored \cite{1996_Ouyang_Flesselles}.  This provides evidence that frozen vortex chimeras can be realized experimentally in much the same fashion as in our numerical procedure.
One important experimental consideration is the impact of imperfect experimental conditions---the basin of attraction of the frozen vortex chimeras must not be inaccessibly small if they are to be found in reality.  To investigate this question, we have randomly perturbed the chimera state and observed its subsequent recovery or destruction (see Supplemental Material, Sec.\,S2 \cite{SM}).  So long as perturbations are not too large, the system is attracted back towards the frozen vortex chimera, thus providing evidence that with sufficiently controlled experimental conditions, frozen vortex chimeras should be experimentally accessible. 

To analyze continuous chimera states, we employ a local coupling field approach similar to the one introduced in Ref. \cite{Kuramoto_Battogtokh:2002}; local order parameters have also recently found application in the synchronization of complex networks \cite{Arenas_Zhou:2008}. Rather than relying on the (discrete) local coupling fields used there, we derive the appropriate local coupling field in the continuous case of Eq.\,\eqref{cgle} by first differentiating $A= r{\mathrm e}^{i\theta}$ to obtain 
\begin{equation}
\frac{d\theta}{dt} = \frac{1}{2i}\left(\frac{1}{A}\frac{{\partial}A}{{\partial}t}-\frac{1}{A^*}\frac{{\partial}A^*}{{\partial}t}\right), 
\label{dtheta}
\end{equation}
where the asterisk denotes complex conjugation. 
Using Eq.\,\eqref{cgle} in Eq.\,\eqref{dtheta}, we can identify the coupling terms as those involving spatial derivatives, namely $(\sqrt{1+c_1^2}/2ir)({\mathrm e}^{i(\alpha-\theta)}\nabla^2A -{\mathrm e}^{-i(\alpha-\theta)}\nabla^2A^* )$, where   $\alpha \equiv \arctan c_1$.
To obtain evolution equations for $r$ and $\theta$ analogous to those for discrete systems,
it follows that the local coupling field 
should be defined as  $R{\mathrm e}^{i\Theta} \equiv (\sqrt{1+c_1^2}/r)\nabla^2A$.  
Indeed, using this coupling field, Eq.\,\eqref{cgle} can be expressed as
\begin{align}
&\frac{d\theta}{dt} = c_3r^2+R\sin(\Theta-\theta+\alpha), \label{mean2} \\
&\frac{dr}{dt} = r(1-r^2)+Rr\cos(\Theta-\theta+\alpha), \label{mean3}
\end{align}
where the main difference from discrete phase oscillators is that the frequency term in the $\theta$ equation is a function of $r$ and the coupling field is differential in $r$ and $\theta$. 

Assuming a coupling field with time-independent $R(x,y)$
and $\Theta(x,y,t)=\Omega{t}+\Phi(x,y)$ with $\Omega$ and $\Phi(x,y)$  time independent,  a coherent solution is one with   $dr/dt= 0$ and   $d\theta/dt=\Omega$.
Noting that $R^2=R^2\sin^2(\Theta-\theta+\alpha)+R^2\cos^2(\Theta-\theta+\alpha)$ and solving Eqs. \eqref{mean2} and\eqref{mean3} for the trigonometric functions, the coherent solutions must satisfy $R^2=(c_3r^2-\Omega)^2+(1-r^2)^2$.  The amplitudes of these coherent solutions are
\begin{equation}
\label{sol}
r = \sqrt{\frac{1+c_3\Omega \pm \sqrt{\left(1+c_3^2\right)R^2-\left(c_3-\Omega\right)^2}}{1+c_3^2}},
\end{equation}
where, given that $R$ is real and positive,  the condition for a (real) solution to exist is $R\ge R_{\mathrm{c}}\equiv (|c_3-\Omega| / \sqrt{1+c_3^2})$.

\begin{figure}[b!]
\includegraphics[width=0.9\columnwidth]{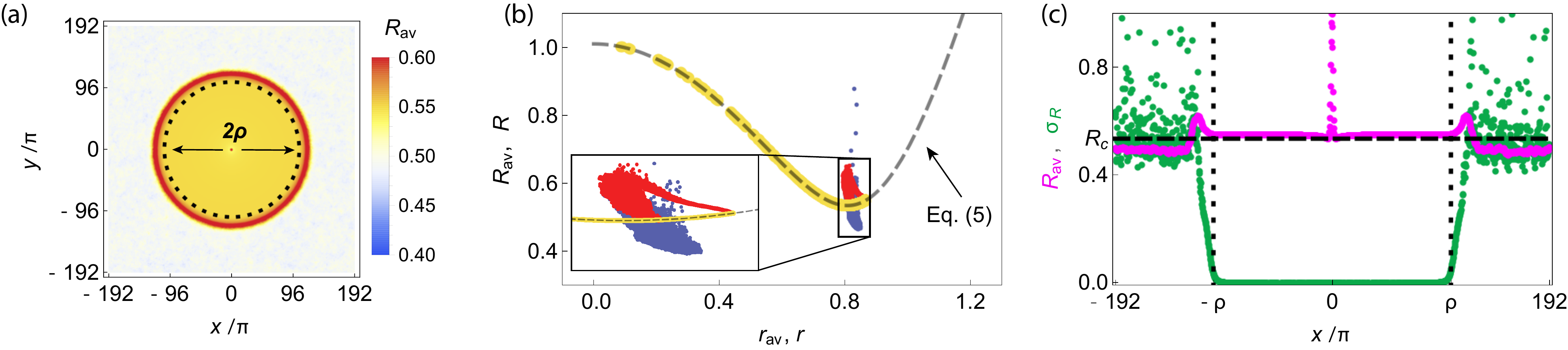}
\caption{
Time-averaged local coupling field $R_{\mathrm{av}}$  for the conditions in Fig.\,\ref{fig2}
as a function of: (a) the two spatial coordinates; (b) the time-averaged amplitude $r_{\mathrm{av}}$ for points in the coherent region (yellow dots), halo region (red dots), and remaining incoherent region (blue dots);
and (c) the $x$ coordinate along the line  $y=0$ (magenta dots).
The inset in (b) magnifies  the box, which includes most points in the incoherent region.  
In (c) we also show the fluctuations $\sigma_R$ of $R$ (green dots).
The dotted lines in (a) and (c) delimitate the coherent domain in Fig.\,\ref{fig2}.
The dashed line in (b) marks the solution in  Eq.\,\eqref{sol} and in (c) marks the critical value $R=R_{\mathrm{c}}$
 above which this (real) solution exists. 
\label{fig3}}
\end{figure}

Figure \ref{fig3}{(a)} shows the time-averaged local coupling field $R_{\mathrm{av}}$ of a frozen vortex chimera.
As with other chimera states considered in the literature, we see that the coupling field amplitude is sufficiently large in the coherent domain to induce synchronization while
it is too small to do so in the incoherent domain (with the exception of the red halo region surrounding the coherent domain).
Figure \ref{fig3}{(b)} shows $R_{\mathrm{av}}$ and the time-averaged amplitude $r_{\mathrm{av}}$,
where it is clear that the solution in Eq.\,{\eqref{sol}} (dashed line) corresponds to the coherent domain (yellow dots).
Note, however, that a portion of the desynchronized domain has
$R_{\mathrm{av}}$  larger (not smaller) than the respective solution in Eq.\,\eqref{sol}, i.e.,  it satisfies 
$R_{\mathrm{av}}> R_{\mathrm{c}}=0.54$.  
The dots marked red in that portion correspond to the red halo surrounding the coherent domain in Fig.\,\ref{fig3}{(a)}. 
To understand why this halo region does not synchronize with the coherent domain,
we must consider the fluctuations of the local coupling field. These fluctuations are
 quantified as the standard deviation $\sigma_R$ calculated over the time series of $R$ and are shown in Fig.\,\ref{fig3}{(c)}.

A distinguishing property of the frozen vortex chimeras 
 apparent in Fig.\,\ref{fig3}{(c)} is that, while negligible in the coherent domain, 
 the fluctuations of the local coupling field rise in the halo region and saturate to large values 
 in the amplitude turbulent domain.
In the discrete nonlocal coupling scenario considered in the original formulation of the self-consistent mean-field
approach \cite{Kuramoto_Battogtokh:2002},  where many oscillators contribute to the 
mean field (in fact all of those for which the coupling kernel is nonzero), fluctuations in the mean-field solution are negligible in both the coherent {\it and} the incoherent domains.
Incidentally, this underlies the increasing stability of chimera states with increasing system size in discrete network systems \cite{Wolfrum_Omelchenko:2011}, rendering the thermodynamic limit of such systems sharply different from the continuous problem considered here.
We argue that the origin of the difference in the nature of the fluctuations derives from the fact that
the CGL system \eqref{cgle}  is continuous and the coupling is local. Thus,  
the portion of the medium contributing to the local coupling field is not large enough to average out the fluctuations.  

We propose that the loss of synchrony across the halo region is driven by these enhanced
fluctuations.  The media in the halo is inclined to synchronize with the spiral because of the large local coupling field, but the large fluctuations present in the amplitude turbulent domain diffuse into the halo region and frequently disrupt this synchronization. A balance is achieved in which the inner spiral is shielded from the fluctuations in the amplitude turbulent domain by the halo, where the fluctuations decay and synchronization is repeatedly achieved and lost.  
To test this mechanism, we performed systematic simulations in which we directly modulate the fluctuations in the amplitude turbulent portion of the media (see Supplemental Material, Sec.\,S3 \cite{SM}).  Increasing the scale of the fluctuations causes the spiral to shrink in size, while decreasing them causes the spiral to grow.   These simulations thus support the proposed fluctuation-based mechanism limiting the growth of the coherent spiral.

In summary, we have studied a novel chimera state appearing in the continuous locally-coupled complex Ginzburg-Landau equation. 
We noted that the nucleation of spiral structures out of an amplitude turbulent domain becomes negligibly small for a range of intermediate values of the parameter $c_3$, and thus that the chimeras can persist without change for very long times.
In contrast to fluctuations in chimera states in nonlocally-coupled discrete systems, the fluctuations in the local coupling field of frozen vortex chimeras cannot be neglected \cite{localcoupling}. 
We conjecture that such fluctuations are responsible for the breakdown of coherence at the boundary between the coherent and incoherent domains.  This appears to reflect a fundamental difference between the chimeras investigated here and those considered previously in nonlocal variants of the CGL equation \cite{Schmidt_Krischer:2015}. This fluctuation-based mechanism provides insights into experimental observations of coexisting 
\onecolumngrid

\twocolumngrid
\newpage \phantom
\newpage \phantom
\newpage \phantom
\newpage
\noindent   order and disorder in continuous fluid media.

\begin{acknowledgments}
The authors acknowledge insightful discussions concerning theoretical aspects of continuous chimera states with Matthias Wolfrum and Jean-R\'egis Angilella and experimental aspects of the BZ system with Oliver Steinbock and Seth Fraden.
This work was supported by ARO Award No.\,W911NF-15-1-0272 and NSF Award No. NSF-CMMI-1435358.
\end{acknowledgments}




\newpage

\setcounter{equation}{0}
\setcounter{figure}{0}
\setcounter{table}{0}
\setcounter{page}{1}
\makeatletter
\renewcommand{\theequation}{S\arabic{equation}}
\renewcommand{\thefigure}{S\arabic{figure}}

\clearpage
\fontsize{11}{2}
\fontfamily{cmr}\selectfont
\baselineskip=1.5em
\setcounter{MaxMatrixCols}{20}
\onecolumngrid

\begin{center}
{\bf\Large {Supplemental Material to ``Chimera States in Continuous Media: Existence and Distinctness''}}\\ 
\vskip 1.0mm

{Zachary G. Nicolaou, Hermann Riecke, and Adilson E. Motter}
\end{center}

\section{Numerical convergence to the continuum limit}
\label{sec1}
To integrate Eq.~(1), we employ a pseudospectral method, which uses a truncated Fourier series that corresponds to a finite set of grid points in space.  As the number of Fourier modes increases, or, equivalently, as the spatial grid spacing decreases, the pseudospectral method converges to the continuum limit with an error that decreases exponentially fast \cite{Boyd2001,Canuto2012}.  
Furthermore, we used an adaptive time step in the integration with a fourth order embedded Runge-Kutta-Feldberg step \cite{1993_Bank_Varga} that achieves a prescribed error tolerance---smaller error tolerances result in smaller time steps being taken by the integrator.  We have tested that our results remain valid and unchanged as the grid spacing and error tolerances are further reduced at fixed system size, and thus that the numerical simulations have converged to the continuum limit of interest.  This truly continuous continuum limit should not be confused with the continuum limit considered in discrete nonlocal coupling scenarios, which (in spite of the name) does not result in a spatially continuous phase field variable.  The spiral radius was the quantity that we observed to converge most slowly to the continuum limit.  Figure S1(a) shows the measured spiral radius $\rho$ of a frozen vortex chimera state as a function of the numerical grid spacing (at fixed error tolerance of $10^{-10}$) for a frozen vortex chimera with $c_1=2.0$ and $c_3=0.85$.   We found that once the grid spacing is much finer than the natural ${\cal{O}}(1)$ length scale of the problem, the numerical results converge.  The instantaneous boundary between the coherent and incoherent domains is a fluctuating quantity, and thus there is some scatter associated with the points in Fig.\ S1(a).  In order to obtain a reliable estimate, the spiral radius is measured by averaging the oscillator frequency over many such fluctuations and determining where this average frequency rises above the spiral's frequency. Through computational tests, we found that a minimum of twenty oscillation periods is needed to determine the spiral radius accurately once the spiral reaches its full size, but many more were used in our simulations.
 \begin{figure}[htb]
\includegraphics[width=1.0\columnwidth]{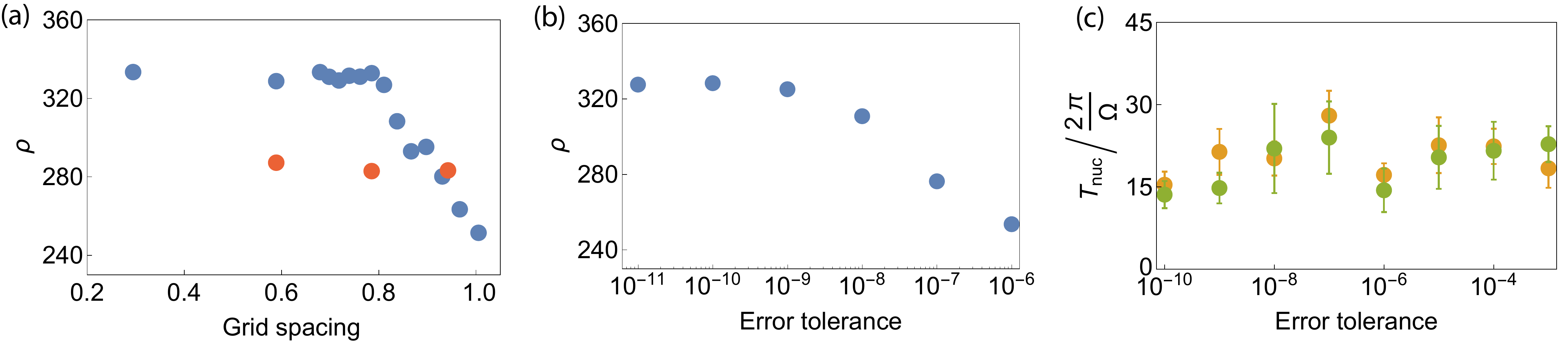} 
\caption{Convergence of numerical results. (a) Spiral radius vs. grid spacing for pseudospectral integration (blue dots) and finite-difference integration (red dots). (b) Spiral radius vs. error tolernace for pseudospectral integration. (c) Mean nucleation time (in a domain of amplitude turbulence) vs. error tolerance for pseudospectral integration with grid spacings of $L/1536 \approx 0.78$ (green dots) and $L/768 \approx 1.57$ (orange dots), and bars showing the estimated sampling error of the mean (for ten independent simulations with random initial conditions).  With sufficiently small grid spacing (a) and error tolerance (b) in the pseudospectral integration, the chimera state properties converge to their continuum limits. In (c), the estimation of the time required for spiral nucleation out of amplitude turbulence does not show any significant trend with grid and error tolerance refinement, which indicates that the results are already converged. The parameters are $c_1=2.0$ and $c_3=0.85$ in (a, b) and  $c_1=2.0$ and $c_3=0.71$ in (c). \label{figs1}}
\end{figure}

Figure S1(b) shows the measured spiral radius for this same chimera state (with fixed grid spacing of $L/1536\approx0.78$) as a function of the error tolerance.  The spiral radius converges to a maximum size for error tolerances below $10^{-9}$, where the chosen time steps become sufficiently small to ensure convergence. For simulations in the main text, we used a grid spacing of $L/1536 \approx 0.78$ with an error tolerance of $10^{-10}$ to ensure that the results are converged.  To confirm that this grid spacing was sufficiently fine, we also performed tests with an error tolerance of $10^{-10}$ and extremely fine grid spacings of $L/2048 \approx 0.59$ and $L/4096 \approx 0.29$, as shown in Fig.\  S1(a). The results of these extremely expensive simulations did not differ significantly from those used in the main text, indicating that those simulations are indeed sufficiently converged.

Other quantities besides the spiral radius, such as the average local coupling field, fluctuations, and spiral nucleation rates, converged already at coarser grid spacing and with larger error tolerances to within the observed sampling error.  For example, the time $T_{\mathrm{nuc}}$ required for spiral nucleation out of amplitude turbulence is shown in Fig.\ S1(c) for $c_1=2.0$ and $c_3=0.71$, where the latter was chosen such that the nucleation time was not too large to be determined numerically.  The estimated sampling error of the mean over the ten samples in the estimate for the spiral nucleation time was larger in Fig.\ S1(c) than any numerical trend with the grid spacing or error tolerance.  Such lower precision simulations were used to estimate the nucleation rates and in very long runs, while high precisions simulations are used everywhere else.  

Finally, to demonstrate that the chimera state is not an artifact of the pseudospectral method, we also employed an entirely different method and implementation.
We used Mathematica's numerical integration function NDSolve with a finite-difference approximation (using a difference order/stencil size of $12$) for the spatial derivatives and Adam's multistep method (using an adaptive time step and adaptive method order).  The estimated spiral radius in these finite-difference simulations are shown as red dots in Fig.\ S1(a). These simulations confirmed the stability of the chimera state. For coarse grid spacing of $L/1280 \approx 0.94$, the measured spiral radius was similar to the pseudospectral measurements with similar grid spacing.  As expected of a finite-difference implementation, its convergence when refining the grid was much slower than for the pseudospectral method, and its low computational efficiency when it comes to high-precision results made a convergence to the pseudospectral result prohibitively expensive. As noted above, albeit to a lesser extent, the pseudospectral simulations also become increasingly expensive as the grid is refined. This is not a problem in our study, however, because we were able to perform all pseudospectral simulations using a grid spacing below the scale required for convergence to  the continuum.

\newpage

\section{Resilience to changes in the initial conditions}
As described in the main text, the initial conditions used to prepare frozen vortex chimeras were obtained from spiral states that were first generated at parameter values of $c_1$ and $c_3$ for which spiral nucleation occurs readily, followed by a quasi-static parameter change.  To observe any chimera state, the initial conditions must be in its basin of attraction, meaning that the system will evolve into that chimera state on a time scale much shorter than the lifetime of the state.  To test the resilience of the frozen vortex chimeras to changes in the initial conditions and to shed some light on the size of the associated basins of attraction, we have applied random perturbation to these states to determine how large a perturbation is required to be in order to destroy the structure.  Figure S2(a) shows the time evolution of the spiral radius  under the influence of such random perturbations.  At time $t=0$, a random perturbation $\delta A(x,y)$ was added to $A$.  The perturbation was chosen with random Fourier modes that decay in amplitude with the same exponential rate as observed in amplitude turbulence so as to have similar length scales to amplitude turbulence [shown in Fig.\ S2(b)].  To measure the size of this perturbation, we use the Sobolev norm $\lVert \delta A \rVert_{\mathrm{Sob}} \equiv \left(\int{\mathrm d}x{\mathrm d}y ~ \left(|\delta A|^2 + |\nabla \delta A|^2\right)/L^2 \right)^{1/2} $ since variation in both the amplitude and its derivative are relevant.  We observe that the initial perturbation destroys the outer portion of the spiral, but the spiral grows back to its original size and form provided the initial perturbation norm is not too large.
\begin{figure}[hbt]
\includegraphics[width=1.0\columnwidth]{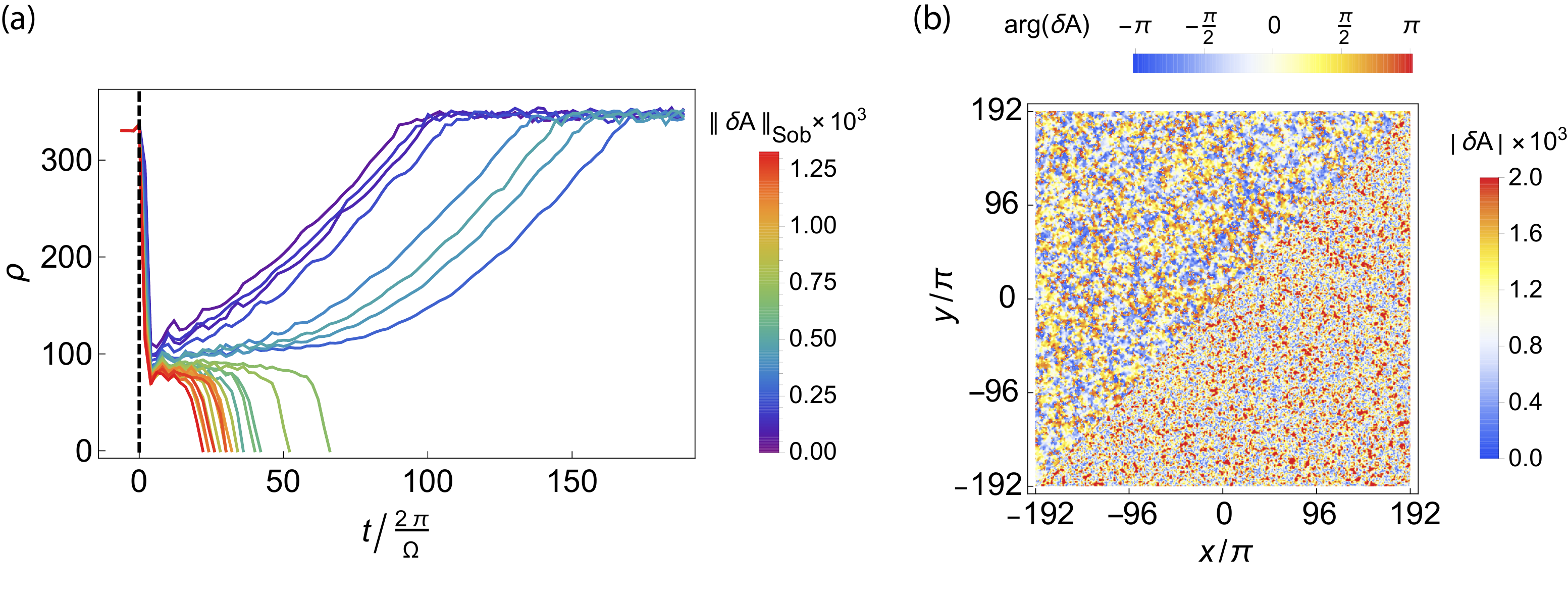} 
\caption{(a) Spiral radius versus time following a random perturbation.  At time $t=0$ (black dashed line), random perturbations of varying magnitude were applied to a frozen vortex chimera, and the subsequent development was tracked.  Shortly after the perturbation, the outer portion of the spiral was destroyed and the spiral radius dropped sharply. For perturbations with norms $\lVert \delta A\rVert_{\mathrm{Sob}}\times 10^{3} < 0.5$, the spiral slowly grew back to its original size, while for larger perturbations, the spiral eventually vanished.   (b) One such random perturbation with norm $\lVert \delta A \rVert_{\mathrm{Sob}} \times 10^3 =  1.3$. \label{figs2}}
\end{figure}
\newpage

\section{Validation of the fluctuation-based mechanism}
To validate the proposed fluctuation based mechanism, which we argue limits the size of the coherent domain, additional simulations have been implemented that attempt to directly modulate the fluctuations present in the amplitude turbulent domain. To accomplish this modulation, an additional diffusive term $\pm \frac{1}{2} \nabla^2 A$ was included in Eq. (1) only for the amplitude turbulent portion of the domain that reduced or enhanced the scale of the fluctuations.  This additional term smoothes out the amplitude turbulence and reduces fluctuations when the sign is positive, but creates small-scale variations and enhances fluctuations when the sign is negative.  By only modifying the media outside the coherent spiral, the properties of the spiral such as its frequency and wavelength are left unaltered, and only the scale of the fluctuations in the surrounding environment is affected. Figure S3(a) shows that distribution of measured fluctuations $\sigma_R$ in the amplitude turbulence region in these simulations; the inset illustrates resulting growth and shrinkage when the fluctuations are reduced and enhanced, respectively.  The mean value of the local coupling field in the amplitude turbulent domain was also affected by the modulation; in fact, it increased when the fluctuations were enhanced and decreased when the fluctuations were reduced, as shown in Fig.\ S3(b).  Naively, an increased local coupling field should induce more synchronization, {\em but the opposite is observed in these simulations}.  Thus these simulations further demonstrate that the local coupling field fluctuations can dominate over the mean field value and inhibit synchronization.
\begin{figure}[hbt]
\includegraphics[width=1.0\columnwidth]{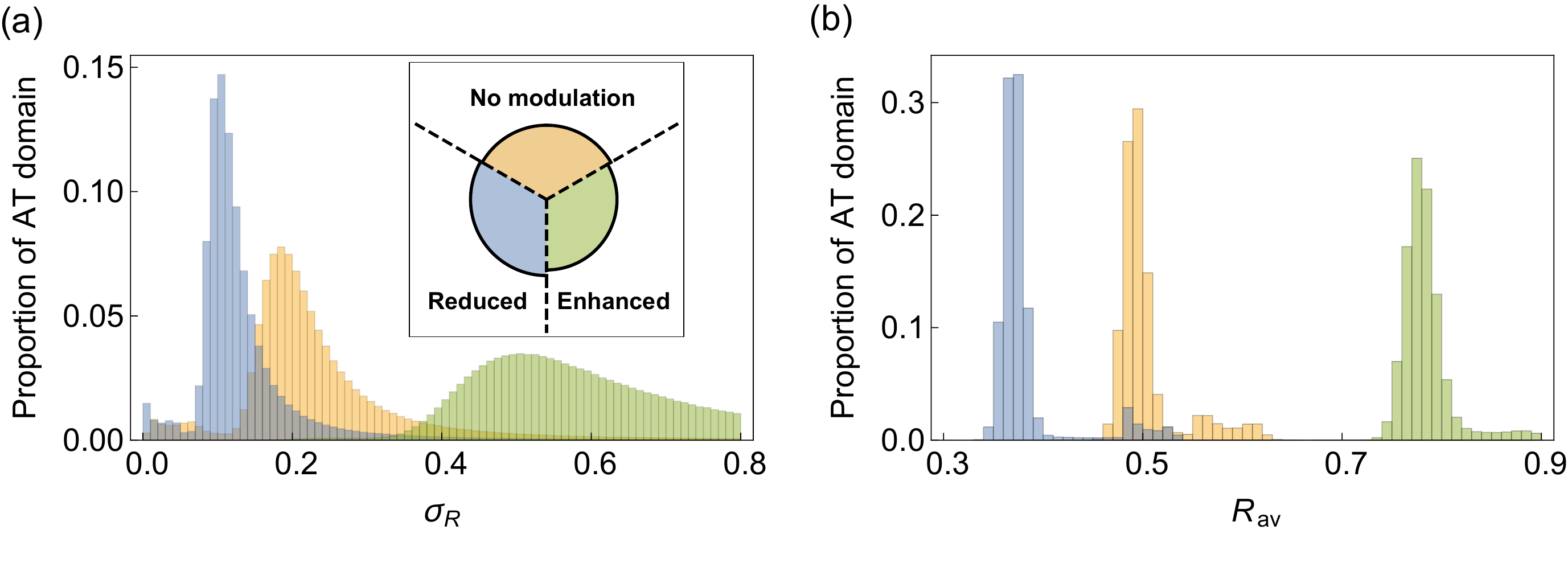}
\caption{Modulation of local coupling field fluctuations. Histograms of the amplitude turbulent (AT) domain with (a) local coupling field fluctuations $\sigma_R$ and (b) time-averaged local coupling field $R_{\mathrm{av}}$, in simulations of frozen vortex chimeras.  These simulations are for $c_1=2.0$ and $c_3=0.85$ with no modulation (orange bars), with reduced fluctuations in the AT domain (blue bars), and with enhanced fluctuations in the AT domain (green bars). The inset in (a) shows measured spiral radius in each case. The spiral was observed to grow by  $2.4\%$ when fluctuations were reduced and shrink by $5.2\%$ when the fluctuations were enhanced.  \label{figs3}}
\end{figure}
\newpage

\section{Animation of a frozen vortex chimera}
An animation of a frozen vortex chimera (with sufficiently small grid spacing and error tolerance to ensure convergence to the continuum limit) has been produced and is available as part of this Supplemental Material.  A snapshot of this animation is shown in Fig.\ S4.  
\begin{figure}[hbt]
\includegraphics[width=0.5\columnwidth]{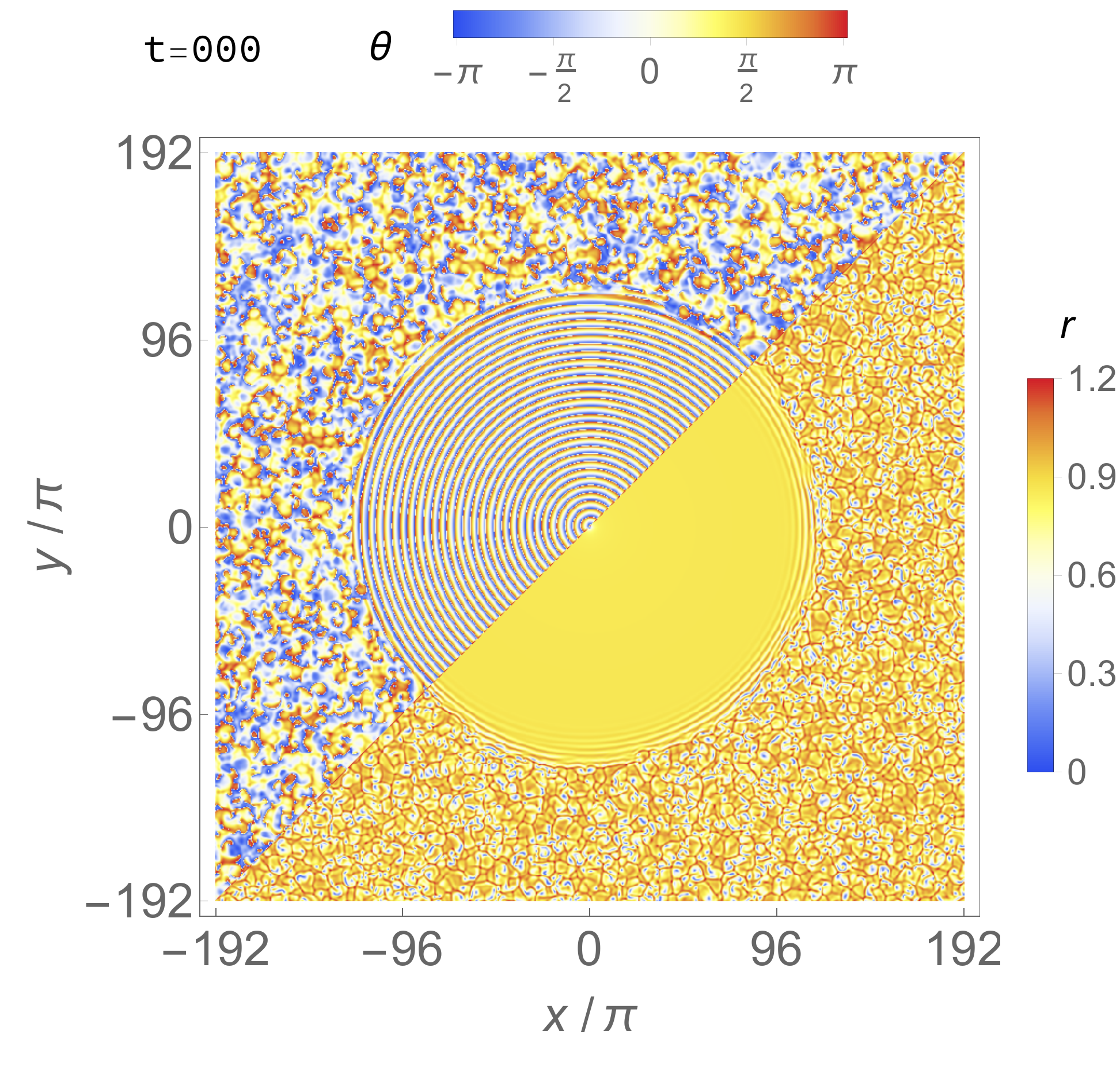}
\caption{Animation of the phase (upper left) and amplitude (lower right) of the field $A=r{\mathrm e}^{i\theta}$ for the frozen vortex chimera with $c_1=2.0$ and $c_3=0.85$.  The central spiral is a portion of coherent media, while the exterior is an incoherent region of amplitude turbulence. The animation starts at time $t=0$ (shown in the upper left) and runs until $t=500$, which is about $12$ spiral oscillation periods $2\pi/\Omega$.  The fit in Fig.~3(b) of the main text suggests that this chimera will persist with no noticeable change (i.e., without spiral nucleation or growth) for over a billion spiral oscillation periods. \label{figs4}}
\end{figure}


\begin{thebibliography}{99}
\bibitem{antispirals} Antispirals are characterized by exhibiting rotation with a phase velocity that points towards the spiral core~\cite{Gong2003}. Here, we omit the prefix and refer to these structures simply as spirals.
\end{thebibliography}
\end{document}